\documentclass{article}
	\usepackage[backend=biber]{biblatex}
	\usepackage{graphicx, float, amsmath, mathtools}
	\title{Minimum Message Length Autoregressive–Moving-Average Model Order Selection}
	\author{Zheng Fang \thanks{Corresponding author. Email: zfan51@student.monash.edu} \thanks{Department of Artificial Intelligence and Data Science, Monash University, Clayton, Victoria 3168, Australia.}, David L. Dowe \thanks{Department of Artificial Intelligence and Data Science, Monash University, Clayton, Victoria 3168, Australia.}, Shelton Peiris \thanks{School of Mathematics and Statistics, University of Sydney, Camperdown, NSW 2006, Australia.}, Dedi Rosadi \thanks{Department of Statistics, Gadjah Mada University, Sleman, Yogyakarta 55500, Indonesia.}}

\begin{document}
	\maketitle
	\begin{abstract}
        This paper derives a Minimum Message Length (MML) criterion for the model selection of the Autoregressive–Moving-Average (ARMA) time series model. The model selection criterion using Wallace and Freeman's (1987) MML approximation, which is an extended version of MML, called MML87. The MML87 performances on the ARMA model compared with other well-known model selection criteria, Akaike’s Information Criterion (AIC), Corrected AIC (AICc), Bayesian Information Criterion (BIC), and Hannan-Quinn (HQ). The experimental results show that the MML87 outperforms the other model selection criteria as it selects most of the models with lower prediction errors and the models selected by MML87 to have a lower mean squared error in different in-sample and out-sample sizes.
        \\
	    \textbf{keywords -- }minimum message length, time series, ARMA, information theory
    \end{abstract}
    
    \section{Introduction}
        Time series data is important nowadays in terms of economics, financial mathematics, weather forecasting, and signal processing. In finance, we have the high-frequency data from the stock market which is presented as time series. In weather forecasting, we have the data of rainfall also presented as time series. It is essential that select a good fitting time series model in given data, which helps to benefits the financial investors to control the risk, social scientists to analyze the population growth. If we are able to select a good fitting time series model, it significantly contributes to the time series community, that's why this topic is popular in the academic and has been researched for the long term.
        
        Minimum Message Length estimator (MML87) is introduced by Wallace and Freeman \cite{wallace1987estimation}, which is the key inductive inference method in this paper, it belongs to the information-theoretic criterion in terms of model selection. MML87 is an extended version of Minimum Message Length (MML) is introduced also by Wallace and Boulton in 1968 \cite{wallace1968information}. MML87 is widely applicable in different areas, and successfully in solving many time series problems, the results from Fitzgibbon and Sak show the MML87 has outperformance in the Autoregressive model (AR) and Moving-Average model (MA) \cite{fitzgibbon2004minimum, sak2005minimum}, and results from Z Fang shows the MML87 outperform in the hybrid ARMA with neural network LSTM (Long short-term memory) \cite{fang2021minimum, fang2021climate}.

        The purpose of this paper is to derive and investigate the use of the MML87 methodology for the Autoregressive–Moving-Average Model (ARMA) model. The ARMA model is one of the most popular time series models introduced by Box and Jenkins in 1976 \cite{box1976time}, because the ARMA model considers the ability in modeling the lag (i.e., past) value in time series, and also modeling the random factors in time series such as strikes, or error residuals. The effects of lag value and lag residuals on the forecasted value will gradually decrease with subsequent periods \cite{chatfield2004cross}. The MML87 in this paper is based on unconditional log-likelihood from the ARMA model. This paper also derives the Fisher information matrix based on the likelihood function, which is the key to make the MML87 to calculable in the ARMA time series case. The MML87 is empirically compared with the AIC, AICc, BIC, and HQ, in terms of the Mean squared error (MSE) in the forecasting window. This paper also compares the number of times that each information-theoretic criterion selects the model with lower mean squared error, the results show that the MML87 is outperforming for the ARMA forecasting than the other information-theoretic criterion. It is because the MML87 selects the model with minimizing the overall message length gives the goodness of fit for the model and less complexity of the model and selected \cite{wallace1987estimation, wallace1999minimum} we will define the message length later in section \ref{mml}.
	
	\section{Time Series Background}
	    \subsection{ARMA Model}
	        This section reviews the theory of Autoregressive Moving AverageModel (ARMA) modelling due to Box and Jenkins (1970) \cite{box2015time}. Let's consider a time series data $\{ y_t \ ; \ t = 1, ..., n\}$ generated from ARMA($p, q$) model:
    		\begin{equation}
    		\label{armatimeseries}
    		    y_t = \phi_1 y_{t-1} + \cdots + \phi_p y_{t-p} + \epsilon_t - \theta_1 \epsilon_{t-1} - \cdots - \theta_p \epsilon_{t-q}
    		\end{equation}
    		where $\{\epsilon_t\} \sim WN(0,\sigma^2)$.  
    		\\
    		\\
    		Rearrange equation \ref{armatimeseries}, the function of white noise $\epsilon_t$ become:
    		\begin{equation}
    		\label{rearmatimeseries}
    		    \epsilon_t = y_t - \sum \limits _{i=1}^p \phi_i B^i y_t + \sum \limits _{i=1}^q \theta_i B^i \epsilon_t
    		\end{equation}
    		\\
    		Denote the $B^i$ is backshift operator where  $B^i y_t = y_{t-i}$ for $i \geq 0$, rearrange the equation \ref{rearmatimeseries}:
    		\begin{equation}
    		\label{backshift}
    		    (1 - \sum \limits _{i=1}^p \phi_i B^i ) y_t = (1 - \sum \limits _{i=1}^q \theta_i B^i) \epsilon_t
    		\end{equation}
    		more generally is $\phi(B) y_t = \theta (B) \epsilon_t$, where $\phi(B) = 1 - \sum \limits _{i=1}^p \phi_i B^i$ and $\theta(B) = 1 - \sum \limits _{i=1}^q \theta_i B^i$.
    		\\
    		\\
    		The $\epsilon_t$ can be rearranged as: 
    		\begin{equation}
    		\label{epsilonbackshift}
    		    \epsilon_t(\beta) = \frac{\phi (B)}{\theta (B)}y_t
    		\end{equation}
    		where $\epsilon_t(\beta)$ is function of $\beta$: $\beta = (\phi_1, ..., \phi_p, \theta_1, ..., \theta_q, \sigma^2)$ such that $\beta$ is a 1 x (p + q + 1) vector.
    		
    		Assuming the data generated from a stationary Gaussian ARMA($p, q$) process, it follows the multivariate Gaussian distribution with zero mean, based on the unconditional log-likelihood function $L(y | \beta)$, N observations $y = (y_1, ..., y_N)$ is:
    		\begin{equation}
    		\label{loglikelihood}
    		    L(y | \beta) = - \frac{N}{2} \log(2 \pi \sigma^2) - \frac{1}{2} \log|\Sigma| - \frac{1}{2 \sigma^2} y^T \Sigma^{-1} y
    		\end{equation}
    		where $\sigma^2 \Sigma$ is n by n autocovariance matrix, $|\Sigma|$ is determinant of $\Sigma$ which is determinant in joint density of $y_t$. The $\Sigma = E(y^T y)$ is n x n theoretical covariance matrix of the data satisfying an ARMA($p, q$) model \cite{fitzgibbon2004minimum, sak2005minimum, anderson1976inverse}.
    	
    	\subsection{Information Theoretic Criterion}
    	    There are some existing model selection techniques from the information theoretic criterion in the ARMA time series model, including the AIC (Akaike's Information Criterion), AICc (Corrected AIC), BIC (Bayesian Information Criterion), and HQ (Hannan-Quinn). The formulas to calculate those information theoretic criterion in ARMA models given as:
    	    \begin{itemize}
    	        \item AIC = $\log{(L)} + \frac{2(q + p + k + 1)}{N}$. 
    	        \\
    	        
    	        \item AICc = $\log{(L)} + \frac{2(p + q + k + 1)}{(N - p - q - k - 2)}$
    	        \\
    	        
    	        \item BIC = $\log{(L)} + \frac{(p + q + k + 1) \log{(L)}}{N}$
    	        \\
    	        
    	        \item HQ = $\log{(L)} + \frac{2(p + q + k + 1) \log{\log{N}}}{N}$
    	        \\
    	        
    	    \end{itemize}
    	    where $k = 1$ if intercept $c \neq 0$ and $k = 0$ if intercept $c = 0$, $N$ is the number of observations for a given time series, and $L$ is likelihood of the data.

	\section{Fisher Information Matrix}
	    In this section, we calculates the Fisher Information matrix required by the MML87, it calculated as negative of the expected value of second derivatives of log-likelihood given as:
	    \begin{equation}
	    \label{fisherformula}
	        -E_{f(X; \beta)} = [\frac{\partial^2}{\partial \beta^2} \log f(X; \beta)]
	    \end{equation}
	    where the $f(X; \theta)$ be a probability density function of $X$ given some parameter $\theta$.
	    
	    The Fisher information matrix for log likelihood of ARMA time series model is combine the equations \ref{loglikelihood} and \ref{fisherformula}: 
		\begin{equation}
		\label{fisher}
		    I(\beta) = - E[ \frac{\partial^2 \log L(y | \beta)}{\partial \beta' \partial \beta}]
		\end{equation}
		\\
		From equation \ref{loglikelihood} and equation \ref{fisher}, the fisher information matrix given as:
		\begin{equation}
		\label{refisher}
		    \frac{\partial^2 \log L(y | \beta)}{\partial \beta' \partial \beta} = - \frac{1}{2} \frac{\partial^2 \log(|\Sigma |)}{\partial \beta' \partial \beta} - \frac{1}{2 \sigma^2} \frac{\partial^2 y^T \Sigma^{-1} y}{\partial \beta' \partial \beta}
		\end{equation}
		\\
		In the meantime, the term
		\begin{center}
		    $- \frac{1}{2} \frac{\partial^2 \log(| \Sigma |)}{\partial \beta' \partial \beta}$
		\end{center}
		from the equation \ref{refisher} will converges to 0 in probability when $n \rightarrow \infty$ then we can ignore this term, because the function is dominated by the second term of equation \ref{refisher} in large values of $n$. 
		
		Based on the Gaussian MLE by maximize equation \ref{loglikelihood}: 
		\begin{center}
		    $\hat \beta = \mathop{\arg\min} \limits _{\theta \in \Theta}[\log \{ \frac{y^T \Sigma^{-1} y}{n} \} + \frac{ \log |\Sigma|}{n} ]$
		    
		    $\hat \sigma^2 = \frac{y^T \Sigma^{-1} y}{n}$
		\end{center}
		So it is important to note that $\hat \beta$ does not depend on $\sigma^2$. The exact least squares estimate to provide the approximations really close to the results from Gaussian MLE \cite{yao2006gaussianii, yao2006gaussiani}.
		\\
		Let 
		\begin{equation}
		\label{Sequation}
		    S(\phi, \theta) = y^T \Sigma^{-1} y
		\end{equation}
		and $\epsilon_t(\beta)$ is the white noise function of $\beta$ with $\epsilon_t(\beta) = \frac{\phi (B)}{\theta (B)}y_t$ according to equation \ref{epsilonbackshift}. 
		\\
		Rearrange the equation \ref{epsilonbackshift} and \ref{Sequation}, which is $\epsilon_t$ in function of $\beta$, the formula given by:
		\begin{equation}
		    S(\phi, \theta) =  \sum \limits _{t=1}^n \epsilon_t^2(\beta)
		\end{equation}
		where the mean of white noise $\epsilon_t$ is 0 \cite{miller1995exact, wincek1986exact, shephard1997relationship}. Box and Jenkins (1976) and Box, Jenkins, and Reinsel (2015, Section 7.1.2-4 p 210-213.)\cite{ljung1979likelihood, box2015time, box1976time} point out the parameters can be obtained by minimize the unconditional sum of squares function: $S(\phi, \theta)$ .
		\\
		The Fisher Information matrix \ref{fisher} can be expressed as only the second term of equation \ref{refisher} inside the expectation:
		\begin{center}
		    $\frac{\partial^2 S(\phi, \theta)}{\partial \beta' \partial \beta} = 2n[\frac{\partial \epsilon_t^T (\beta)}{\partial \beta'}  \frac{\partial \epsilon_t (\beta)}{\partial \beta} \ + \ \epsilon_t (\beta) \frac{\partial^2 \epsilon_t (\beta)}{\partial \beta' \partial \beta}]$
		\end{center}
		by using the product rule. The term 
		\begin{center}
		    $[\frac{\partial \epsilon_t^T (\beta)}{\partial \beta'}  \frac{\partial \epsilon_t (\beta)}{\partial \beta} \ + \ \epsilon_t (\beta) \frac{\partial^2 \epsilon_t (\beta)}{\partial \beta' \partial \beta}]$
		\end{center}
		can be rearranged to 
		\begin{center}
		    $[\frac{\partial \epsilon_t (B)}{\partial \beta'} \frac{\epsilon_t (\beta)}{\partial \beta}]$
		\end{center}
		by using the product rule again \cite{vahid1999partial}.
		\\
		\\
		So the Fisher Information Matrix will be rearranged as:
		\begin{equation}
		\label{fisherfinal}
		    I(\beta) = \frac{n}{\sigma^2} E[\frac{\partial \epsilon_t(\beta)}{\partial \beta'} \frac{\partial \epsilon_t (\beta)}{\partial \beta}]
		\end{equation}
		In the meantime, the Fisher information matrix can be written in a general format regarding $\phi$ and $\theta$:
		\begin{center}
		    $I(\beta) = \begin{pmatrix} I_{\phi \phi} & I_{\phi \theta} \cr I_{\theta \phi} & I_{\theta \theta} \end{pmatrix}$
		\end{center}
		Based on $\epsilon_t(\beta) = \frac{\phi (B)}{\theta (B)}y_t$ from equation \ref{epsilonbackshift} and the lag $k$ of $\epsilon_t$ is $\frac{B^k \phi (B)}{\theta (B)} y_t$ \cite{klein1995computation}.
		\begin{center}
		    $\frac{\partial \epsilon_t(\beta)}{\partial \phi_k} = \frac{\partial \phi (B) \theta^{-1} (B) y_t}{\partial \phi_k} = - B^k \theta^{-1} (B)y_t = - \phi^{-1} (B) \epsilon_{t-k}$
		\end{center}
		\begin{center}
		    $\frac{\partial \epsilon_t(\beta)}{\partial \theta_j} = \frac{\partial \theta^{-1} (B) \phi (B) y_t}{\partial \theta_j} = -\theta^{-2} (B) (-B^j)\phi (B)y_t = \theta^{-1}(B)\epsilon_{t-j}$
		\end{center}
		Consequently the element $(k, \ j)$ of the $(\phi, \ \theta)$ block of the Fisher information matrix is:
		\begin{equation}
		\label{fisherforMML}
		    I_{\phi \theta}(\beta) = \frac{n}{\sigma^2}E[\{-\phi^{-1} (B)\epsilon_{t-k}\}\{\theta^{-1} (B) \epsilon_{t-j}\}]
		\end{equation}
		
	\section{Minimum Message Length}
	\label{mml}
	    Minimum Message Length (MML) is an inductive inference method for the model selection, the MML found on top of the coding theory. The underlying concept for the MML principle is based on the data compression theory. Assuming there is a sequence of time series data, and a set of candidates ARMA models, the MML is assuming that the data is encoded into two parts of messages transmitted from sender to receiver \cite{wallace1968information, wong2019minimum}:
	    \begin{itemize}
	        \item ARMA time series model $\beta \in \beta^{*} \subset \boldsymbol{\beta}$
	        
	        \item Time series data fitted on the ARMA model in the first message component $f(X|\beta)$
	        
	    \end{itemize}
	    where $\boldsymbol{\beta}$ is the parameter space of the ARMA model $f(X)$, and the $\beta^{*}$ is a countable subset of parameter space containing all possible MML estimates parameter $\hat \beta$ \cite{wong2019minimum}.
	    
	    The first dot point in MML theory encodes the model structure and parameters then transmits from sender to receiver, which is called the \textbf{assertion}. The second dot point in MML theory encodes the observed data $X$ using the model specified in the \textbf{assertion}, which is called \textbf{detail}. The MML minimize the total message length transmitted from sender to receiver contains data $X$ and model $\beta \in \beta^{*}$ given as \cite{wallace1968information}:
	    \begin{equation}
	    \label{MML}
	        I(X, \beta) = \underbrace{I(\beta)}_\text{assertion} + \underbrace{I(X | \beta)}_\text{detail}
	    \end{equation}
	    where the $I()$ is represented as the message length transmitted from sender to receiver. The assertion $I(\beta)$ represents the complexity of the model, and detail $I(X | \beta)$ represents how well the model fitting the data. So the MML minimize the tradeoff between the complexity and goodness-of-fit of the model, that's why the MML has fewer chances to overfitting the data.
	    
		The MML87 is extended version of Minimum Message Length (MML) used in model selection of several continuous parameter set $\beta = (\phi_1, ..., \phi_p, \theta_1, ..., \theta_q, \sigma^2)$ \cite{wallace1999minimum}:
		\begin{equation}
		\label{MML87first}
		    I(X, \beta) = \underbrace{- \log \pi(\beta) + \frac{1}{2} \log |F(\beta)| + \frac{k}{2} \log \kappa_k}_\text{assertion} + \underbrace{\frac{k}{2} - \log f(X | \beta)}_\text{detail}
		\end{equation}
		\\
		Rearranging the equation \ref{MML87first} will become:
		\begin{equation}
		    MessLen(\beta, y) = - \log(\frac{\pi(\beta)f(y_1, ..., y_n | \beta) \epsilon^n}{\sqrt{|F(\beta)|}}) + \frac{k}{2}(1 + \log\kappa_k) - \log h(k)
		\end{equation}
		where $\epsilon^n$ is measuring the accuracy of data, $h(k)$ is the prior on the number of parameters, $\pi(\beta)$ is the Bayesian prior distribution over the parameter set, $f(y_1, ..., y_n | \beta)$ is standard statistical likelihood function, $F(\beta)$ is the Fisher Information matrix of the parameter set $\beta, \ \kappa_k$ is the lattice constant (which accounts for the expected error in the log-likelihood function from ARMA model in equation \ref{loglikelihood} due to quantization of the n-dimensional space, it is bounded above by $\frac{1}{12}$ and bounded below by $\frac{1}{2 \pi e}$. 
		\\
		
		For example, $\kappa_1 = \frac{1}{12}$, $\kappa_2 = \frac{5}{36 \sqrt{3}}$, $\kappa_3 = \frac{19}{192 * 2^{1/3}}$, and $\kappa_k \rightarrow \frac{1}{2 \pi e}$ when $k \rightarrow \infty$). Assuming the exact likelihood function is used, and assuming that the data comes from a stationary \cite{fitzgibbon2004minimum, wallace1999minimum, ward2008review}. 
		
		The stationary model message length is calculated by substitution of equation \ref{fisherforMML} and likelihood function into (5) alone with the prior \cite{fitzgibbon2004minimum, sak2005minimum, wallace1999minimum}.
		
	\section{Bayesian Priors}
		In the MML87, it is ignorant prior to observation of data. The MML87 for ARMA model in this report using exact likelihood function and expect the data comes from a stationary process, so placing a uniform prior on $p, q, \phi_1, ..., \phi_p, \theta_1, ..., \theta_q$ and $\log{\sigma^2}$ from the results \cite{fitzgibbon2004minimum, sak2005minimum}.
		\\
		\begin{equation}
		    h(p) \ \propto \ 1
		\end{equation}
		\\
		\begin{equation}
		    h(q) \ \propto \ 1
		\end{equation}
		\\
		\begin{equation}
		    h(\phi_1, ..., \phi_p, \theta_1, ..., \theta_q, \sigma^2) \ \propto \ \frac{1}{R_p} \times \frac{1}{R_q} \times \frac{1}{\sigma^2} = \frac{1}{R_p R_q \sigma^2}
		\end{equation}
		\\
		where $R_p, R_q$ is hypervolume of the stationarity region and based on the results from \cite{barndorff1973parametrization, piccolo1982size}. We can recursively calculate the invertibility region hypervolume by:
		\\
		
		$M_1$ = 2
		\\
		
		$M_{p+1} = \frac{p}{p+1} M_{p-1}$
		\\
		
		$M_{q+1} = \frac{q}{q+1} M_{q-1}$
		\\
		
		$R_p = (M_1M_3 \  \times \ ... \ \times \ M_{p-1})^2$
		\\
		
		$R_q = (M_1M_3 \  \times \ ... \ \times \ M_{q-1})^2$
		\\
		
		$R_{p+1} = R_p M_{p+1}$
		\\
		
		$R_{q+1} = R_q M_{q+1}$
		
	\section{Empirical Comparison}
		\subsection{Simulated data ARMA(1, 1)}
			We use the simulated data to compare the performances of MML87 with AIC, AICc, BIC, and HQ. In this case, using uniform distribution to generate 5 different parameter values for $\phi$ and 2 different parameter values for $\theta$. In each combination of $\theta$ and $\phi$, we generate the different sizes of in-sample and out-sample datasets. We totally have 1,000 different data in each combination of $\phi_1, \phi_2, ..., \phi_5$ and $\theta_1, \theta_2$. For each data set, we use different combination of lag order from 1 to 5 for $p$ and from 0 to 5 for $q$ in ARMA$(p, q)$ to construct the ARMA model, it provides 30 different models as the comparison in one dataset. Then use each MML87, AIC, AICc, BIC, and HQ to select one best model out of 30 models and calculate the lost function, the lost function used in the empirical comparison is MSPE$(p)$:
			\begin{equation}
			    MSPE(p) = \frac{1}{T} \sum \limits _{i = T + 1}^{2T} (y_i - (\hat \phi_1 y_{i-1} +  ... + \hat \phi_p y_{i-p} + \epsilon_i +  \hat \theta_1 \epsilon_{i-1} + ... + \hat \theta_q \epsilon_{i-q}))^2
			\end{equation}
			
			Table \ref{countingcompare} use the in-sample size N = 100, and forecast method used in this case is rolling forecast with a window size of T = 10. Under one data set, we select the prediction errors for minimum values of AIC, AICc, BIC, HQ, and MML87, because the minimum of those information criteria values is the model selected out of 30 different models. Next, we compared the different prediction errors selected from the model selected by AIC, AICc, BIC, HQ, and MML87, the selected model with the smallest prediction error, the information criteria is more efficient. In each combination of simulated parameter combination of $\phi$ and $\theta$, there are 100 data set as mentioned above, we compared the number of times that the AIC, AICc, BIC, HQ, and MML87 can select the model with minimum prediction error.
			\begin{table}[H]
			    \centering
			    \begin{tabular}{|c|c|c|c|c|c|}
			        \hline
					 & MML87 & AIC & AICc & BIC & HQ \\ 
					\hline
					$\phi_1, \theta_1$ & \textbf{45} & 25 & 34 & 41 & 40 \\
					\hline
					$\phi_1, \theta_2$ & \textbf{69} & 62 & 60 & 62 & 68 \\
					\hline
					$\phi_2, \theta_1$ & 42 & \textbf{44} & 40 & 40 & 41 \\
					\hline
					$\phi_2, \theta_2$ & \textbf{75} & 64 & 68 & 72 & 72 \\
					\hline
					$\phi_3, \theta_1$ & \textbf{46} & 36 & 40 & 44 & 38 \\
					\hline
					$\phi_3, \theta_2$ & \textbf{69} & 51 & 57 & \textbf{69} & \textbf{69} \\
					\hline
					$\phi_4, \theta_1$ & \textbf{42} & 32 & 37 & \textbf{42} & 38 \\
					\hline
					$\phi_4, \theta_2$ & \textbf{64} & 51 & 50 & 66 & 57 \\
					\hline
					$\phi_5, \theta_1$ & 46 & 41 & 39 & \textbf{56} & 47 \\
					\hline
					$\phi_5, \theta_2$ & 63 & 69 & 66 & \textbf{72} & 68 \\
					\hline
			    \end{tabular}
			    \caption{\centering Number of times that information-theoretic selects the minimum forecast error in N = 100 T = 10}
			    \label{countingcompare}
			\end{table}
			
			For example, the 1st row table \ref{countingcompare} indicate that using the particular values of parameter combination of $\phi_1$ and $\theta_1$ to generate 100 dataset, there are N = 100 plus T = 10 data points in each dataset. We find out the number of times that MML87 select minimum prediction error is 45 out of 100 for modeling each dataset. There is possible that the summation of each number of times is greater than 100 because there is possible that the two or more information criteria selected the same model for minimum prediction errors. According to the above table \ref{countingcompare}, the MML87 select the most of minimum prediction error models in different combinations of parameters. It significantly demonstrates that the MML87 is outperformance in the model selection for the ARMA time series model compared with AIC, AICc, BIC, and HQ.
			
			The Table from \ref{countingcompare50} to \ref{countingcompare500} shows the comparison of counting number of times that those information theoretical criteria to select the minimum prediction error in N = 50, 150, 200, 300, and 500. The results suggest that the MML87 is able to select more times of lower error model than other information criteria.
			
			Table \ref{MSPEN100T10} show the average MSPE in 100 datasets and different simulated parameters when in-sample size N = 50 and 100 and out-sample size T = 10. In each row, the average of MSPE from the model is selected by the different information criteria. The lower forecast error indicates the better model, it highlighted as bold text, the MML87 outperform other information criteria in the majority of the cases.
			\begin{table}[H]
			    \centering
				\begin{tabular}{|c|c|c|c|c|c|}
					\hline
					 & MML87 & AIC & AICc & BIC & HQ \\ 
					\hline
					$\phi_1, \theta_1$ & \multicolumn{1}{|p{1.5cm}|}{\centering \textbf{1.424699} \\ (0.90860)} & \multicolumn{1}{|p{1.5cm}|}{\centering 1.521890 \\ (0.99921)} & \multicolumn{1}{|p{1.5cm}|}{\centering 1.499588 \\ (1.02887)} & \multicolumn{1}{|p{1.5cm}|}{\centering 1.428254 \\ (0.90914)} & \multicolumn{1}{|p{1.5cm}|}{\centering 1.462676 \\ (0.99725)} \\
					\hline
					$\phi_1, \theta_2$ & \multicolumn{1}{|p{1.5cm}|}{\centering 4.397397 \\ (3.48977)} & \multicolumn{1}{|p{1.5cm}|}{\centering 4.541932 \\ (3.64444)} & \multicolumn{1}{|p{1.5cm}|}{\centering 4.515208 \\ (3.60842)} & \multicolumn{1}{|p{1.5cm}|}{\centering 4.417018 \\ (3.49456)} & \multicolumn{1}{|p{1.5cm}|}{\centering \textbf{4.335861} \\ (3.43196)} \\
					\hline
					$\phi_2, \theta_1$ & \multicolumn{1}{|p{1.5cm}|}{\centering \textbf{1.192377} \\ (0.63315)} & \multicolumn{1}{|p{1.5cm}|}{\centering 1.237391 \\ (0.66865)} & \multicolumn{1}{|p{1.5cm}|}{\centering 1.220480 \\ (0.66442)} & \multicolumn{1}{|p{1.5cm}|}{\centering 1.199164 \\ (0.66119)} & \multicolumn{1}{|p{1.5cm}|}{\centering 1.197432 \\ (0.66071)} \\
					\hline
					$\phi_2, \theta_2$ & \multicolumn{1}{|p{1.5cm}|}{\centering 1.400034 \\ (0.63992)} & \multicolumn{1}{|p{1.5cm}|}{\centering 1.427151 \\ (0.63984)} & \multicolumn{1}{|p{1.5cm}|}{\centering 1.413369 \\ (0.64121)}& \multicolumn{1}{|p{1.5cm}|}{\centering \textbf{1.389479} \\ (0.64031)}& \multicolumn{1}{|p{1.5cm}|}{\centering 1.398019 \\ (0.62572)} \\
					\hline
					$\phi_3, \theta_1$ & \multicolumn{1}{|p{1.5cm}|}{\centering \textbf{1.238596} \\ (0.63055)}& \multicolumn{1}{|p{1.5cm}|}{\centering 1.283056 \\ (0.64058)} & \multicolumn{1}{|p{1.5cm}|}{\centering 1.268670 \\ (0.63659)} & \multicolumn{1}{|p{1.5cm}|}{\centering 1.404347 \\ (0.61788)} & \multicolumn{1}{|p{1.5cm}|}{\centering 1.252059 \\ (0.62601)} \\
					\hline
					$\phi_3, \theta_2$ & \multicolumn{1}{|p{1.5cm}|}{\centering 1.430728 \\ (0.68179)} & \multicolumn{1}{|p{1.5cm}|}{\centering 1.534319 \\ (0.73082)} & \multicolumn{1}{|p{1.5cm}|}{\centering 1.504423 \\ (0.72636)} & \multicolumn{1}{|p{1.5cm}|}{\centering \textbf{1.426555} \\ (0.68741)} & \multicolumn{1}{|p{1.5cm}|}{\centering 1.437177 \\ (0.68837)} \\
					\hline
					$\phi_4, \theta_1$ & \multicolumn{1}{|p{1.5cm}|}{\centering 1.286106 \\ (0.69430)} & \multicolumn{1}{|p{1.5cm}|}{\centering 1.338531 \\ (0.70770)} & \multicolumn{1}{|p{1.5cm}|}{\centering 1.311318 \\ (0.68228)} & \multicolumn{1}{|p{1.5cm}|}{\centering \textbf{1.276376} \\ (0.67433)} & \multicolumn{1}{|p{1.5cm}|}{\centering 1.287715 \\ (0.68372)} \\
					\hline
					$\phi_4, \theta_2$ & \multicolumn{1}{|p{1.5cm}|}{\centering \textbf{1.467448} \\ (0.75440)} & \multicolumn{1}{|p{1.5cm}|}{\centering 1.547590 \\ (0.84311)} & \multicolumn{1}{|p{1.5cm}|}{\centering 1.537011 \\ (0.83362)} & \multicolumn{1}{|p{1.5cm}|}{\centering 1.472064 \\ (0.75156)} & \multicolumn{1}{|p{1.5cm}|}{\centering 1.488488 \\ (0.78266)} \\
					\hline
					$\phi_5, \theta_1$ & \multicolumn{1}{|p{1.5cm}|}{\centering \textbf{2.261384} \\ (1.87727)} & \multicolumn{1}{|p{1.5cm}|}{\centering 2.417009 \\ (1.97325)} & \multicolumn{1}{|p{1.5cm}|}{\centering 2.380226 \\ (1.95017)} & \multicolumn{1}{|p{1.5cm}|}{\centering 2.301430 \\ (1.93637)} & \multicolumn{1}{|p{1.5cm}|}{\centering 2.338255 \\ (1.97315)} \\
					\hline
					$\phi_5, \theta_2$ & \multicolumn{1}{|p{1.5cm}|}{\centering \textbf{1.041781} \\ (0.47548)} & \multicolumn{1}{|p{1.5cm}|}{\centering 1.072443 \\ (0.48750)} & \multicolumn{1}{|p{1.5cm}|}{\centering 1.067345 \\ (0.48744)} & \multicolumn{1}{|p{1.5cm}|}{\centering 1.044731 \\ (0.47907)} & \multicolumn{1}{|p{1.5cm}|}{\centering 1.047712 \\ (0.47553)} \\
					\hline
				\end{tabular}
				\caption{\centering Average of MSPE for the model selected by different information-theoretic in N = 100 T = 10}
				\label{MSPEN100T10}
			\end{table}
			
			Table \ref{MSPEcompareN} test the average of prediction error in out-sample size T = 10 by different amount in-sample size N. It uses the in-sample data size of N = 50, 100, 150, 200, 300, and 500 to compare the average MSPE from the models selected by different information-theoretic criteria. The MML87 outperforms the AIC, AICc, BIC, and HQ in the size N = 50, 100, 150, and 300 because the AIC and BIC tend to overfit the data in the smaller in-sample size.
			\begin{table}[H]
			    \centering
				\begin{tabular}{|c|c|c|c|c|c|}
					\hline
					 & MML87 & AIC & AICc & BIC & HQ \\ 
					\hline
					N = 50  & \textbf{1.796509} & 1.886955 & 1.834129 & 1.801342 & 1.826132 \\
					\hline
					N = 100 & \textbf{1.714055} & 1.792131 & 1.771764 & 1.716704 & 1.724539 \\
					\hline
					N = 150 & \textbf{1.748152} & 1.810529 & 1.809941 & 1.749774 & 1.776628 \\
					\hline
					N = 200 & 1.771909 & 1.833165 & 1.821335 & \textbf{1.764313} & 1.789190 \\
					\hline
					N = 300 & \textbf{1.703449} & 1.741882 & 1.737454 & 1.711380 & 1.712863 \\
					\hline
					N = 500 & 1.719375 & 1.736897 & 1.735968 & \textbf{1.717373} & 1.718875 \\
					\hline
				\end{tabular}
				\caption{\centering Average of MSPE for different in-sample size N}
				\label{MSPEcompareN}
			\end{table}
			
			Table \ref{MSPEcompareT} compares the average of MSPE in the out-sample size T = 10, 30, 50, and 100. The MML87 outperforms other information-theoretic criteria in T = 10, 30, and 50.  
			\begin{table}[H]
			    \centering
				\begin{tabular}{|c|c|c|c|c|c|}
					\hline
					Criterions & MML87 & AIC & AICc & BIC & HQ \\ 
					\hline
					T = 10 & \textbf{1.714055} & 1.792131 & 1.771764 & 1.716704 & 1.724539 \\
					\hline
					T = 30 & \textbf{1.793344} & 1.823455 & 1.810748 & 1.804022 & 1.801981 \\
					\hline
					T = 50 & \textbf{1.901877} & 1.925961 & 1.922792 & 1.914095 & 1.909096 \\
					\hline
					T = 100 & 1.898179 & 1.925083 & 1.917025 & \textbf{1.897318} & 1.903784 \\
					\hline
				\end{tabular}
				\caption{\centering Average of MSPE for different out-sample size T}
				\label{MSPEcompareT}
			\end{table}
		
		\subsection{Simulated data ARMA(p, q)}
			This subsection also uses uniform distribution to simulate the dataset by the different amounts and different values of $\phi_p$ and $\theta_q$. This section compares the MSPE from five information-theoretic criteria in terms of forecast windows used T = 1, 3, 5, 10, 30, 50, 70, 100, 130, 150 and N = 100. Each different forecast window contains and generates 100 datasets, the results show the average of MSPE. Table \ref{MSPEcompareARMA} shows the results from simulated data by ARMA($2, 2)$ to ARMA($5, 2)$.
			\begin{table}[H]
			    \centering
				\begin{tabular}{|c|c|c|c|c|c|}
					\hline
					 & MML87 & AIC & AICc & BIC & HQ \\ 
					\hline
					ARMA(2, 1) & \textbf{1.919648} & 1.983033 & 1.958249 & 1.922295 & 1.949766 \\
					\hline
					ARMA(2, 2) & \textbf{2.051329} & 2.118476 & 2.104302 & 2.053353 & 2.081951 \\
					\hline
					ARMA(3, 1) & \textbf{1.487463} & 1.527326 & 1.521105 & 1.490115 & 1.500657 \\
					\hline
					ARMA(3, 2) & 1.712706 & 1.738684 & 1.721419 & \textbf{1.699448} & 1.712409 \\
					\hline
					ARMA(4, 1) & \textbf{1.961224} & 2.034625 & 2.018761 & 1.978715 & 1.994742 \\
					\hline
					ARMA(4, 2) & \textbf{1.520506} & 1.567478 & 1.556770 & 1.521140 & 1.535147 \\
					\hline
					ARMA(5, 1) & \textbf{2.015772} & 2.084429 & 2.055148 & 2.022353 & 2.034267\\
					\hline
					ARMA(5, 2) & \textbf{1.598522} & 1.647486 & 1.637956 & 1.598533 & 1.620931 \\
					\hline
				\end{tabular}
				\caption{\centering Average of MSPE for different ARMA($p, q$)}
				\label{MSPEcompareARMA}
			\end{table}
			
\begin{figure}[H]
				\centerline{\includegraphics[width=4.88in, height=3.00in]{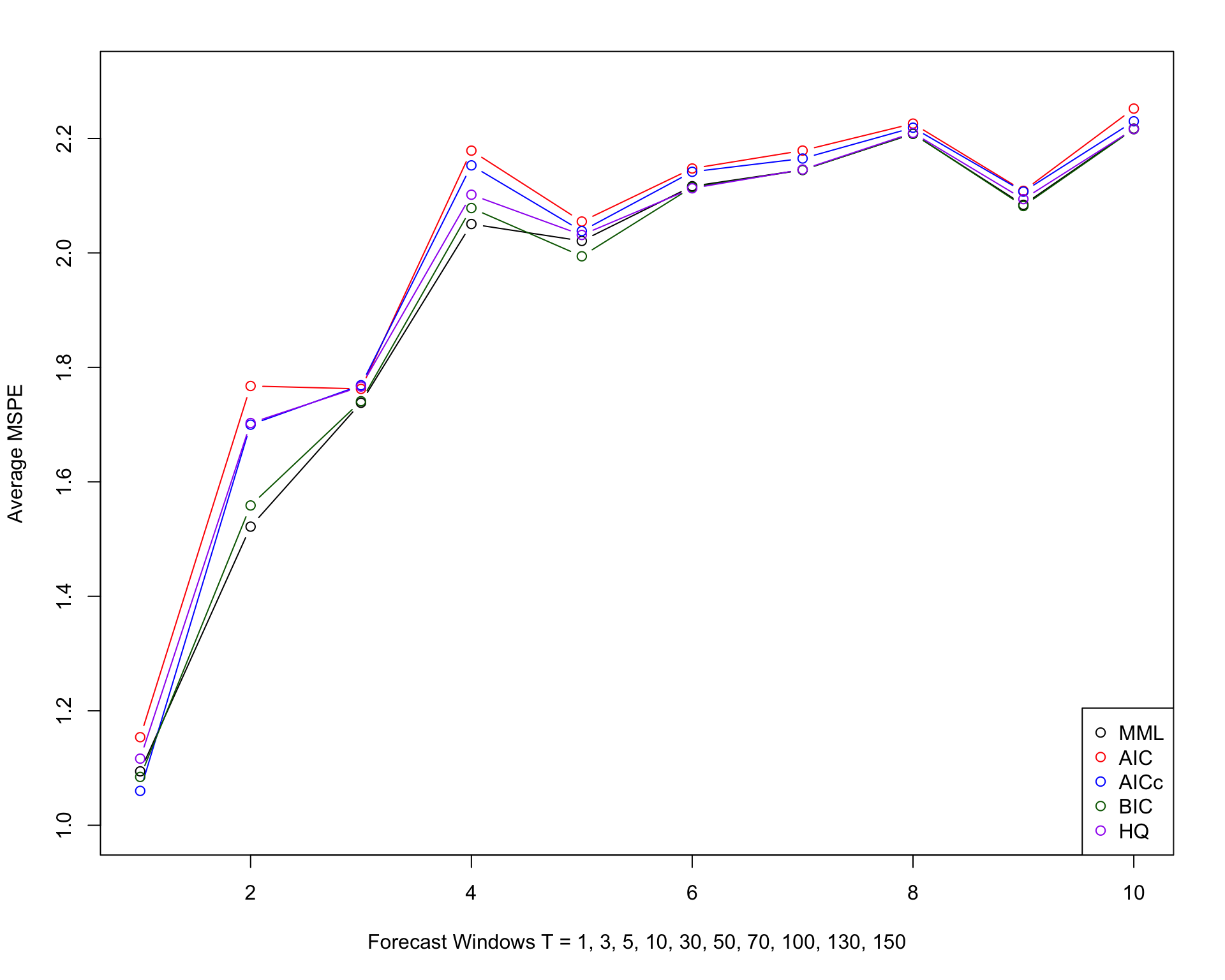}}
					\caption{Average MSPE in ARMA(2, 1) Simulated Data}
					\label{arma21}
			\end{figure}

\begin{figure}[H]
				\includegraphics[width=4.88in, height=2.88in]{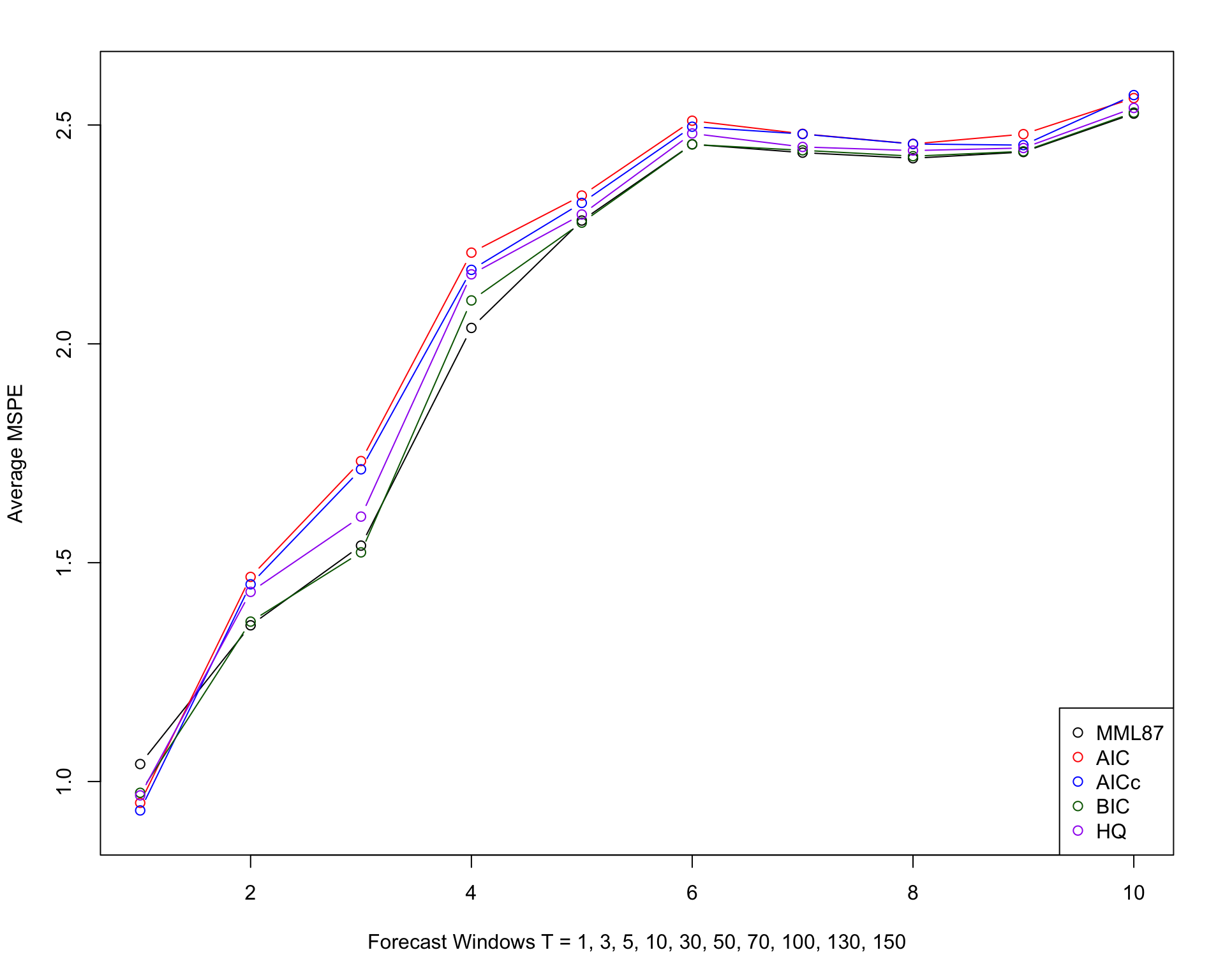}
					\caption{Average MSPE in ARMA(2, 2) Simulated Data}
					\label{arma22}
			\end{figure}

\begin{figure}[H]
				\includegraphics[width=4.88in, height=2.88in]{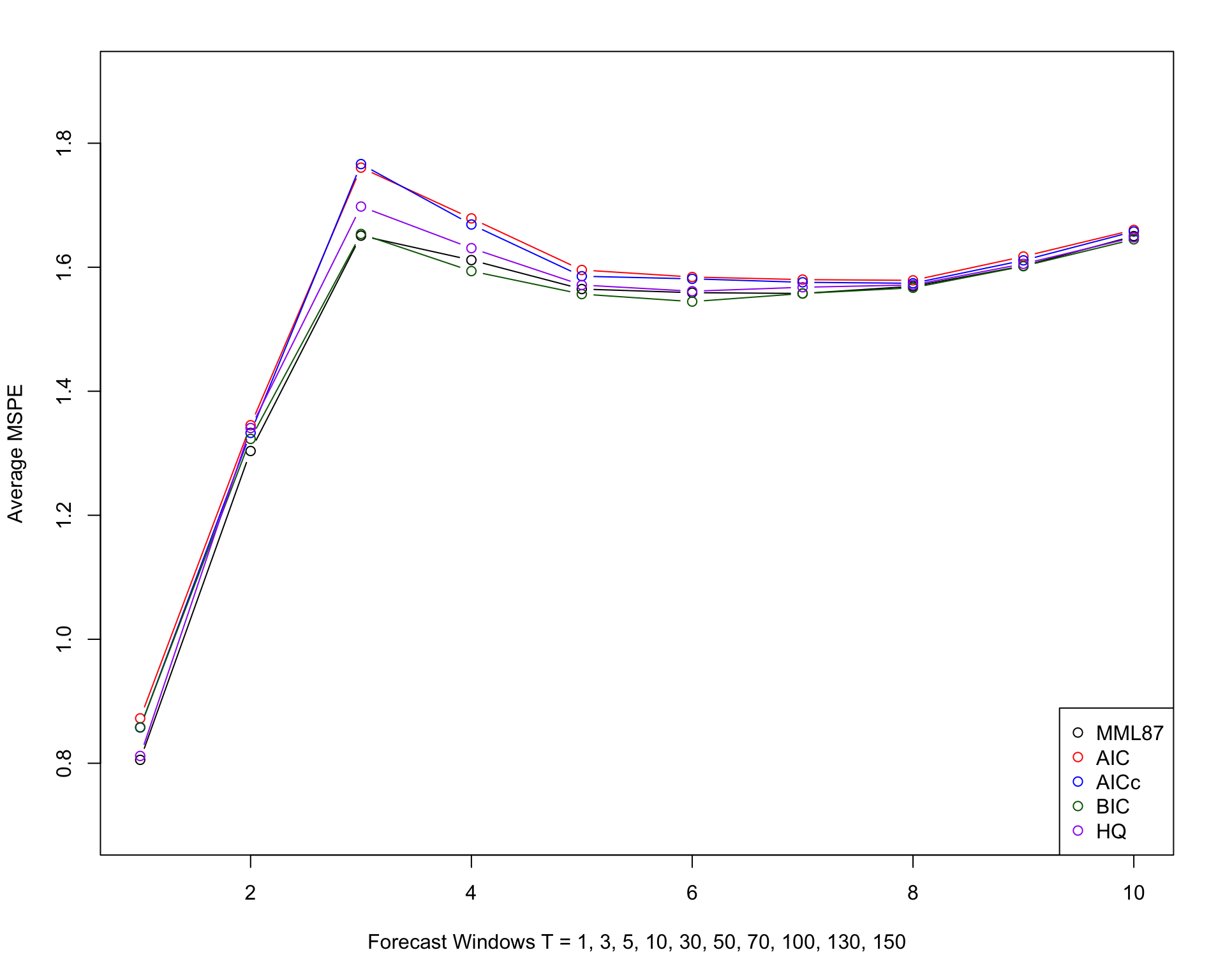}
					\caption{Average MSPE in ARMA(3, 1) Simulated Data}
					\label{arma31}
			\end{figure}

\begin{figure}[H]
				\includegraphics[width=4.88in, height=2.88in]{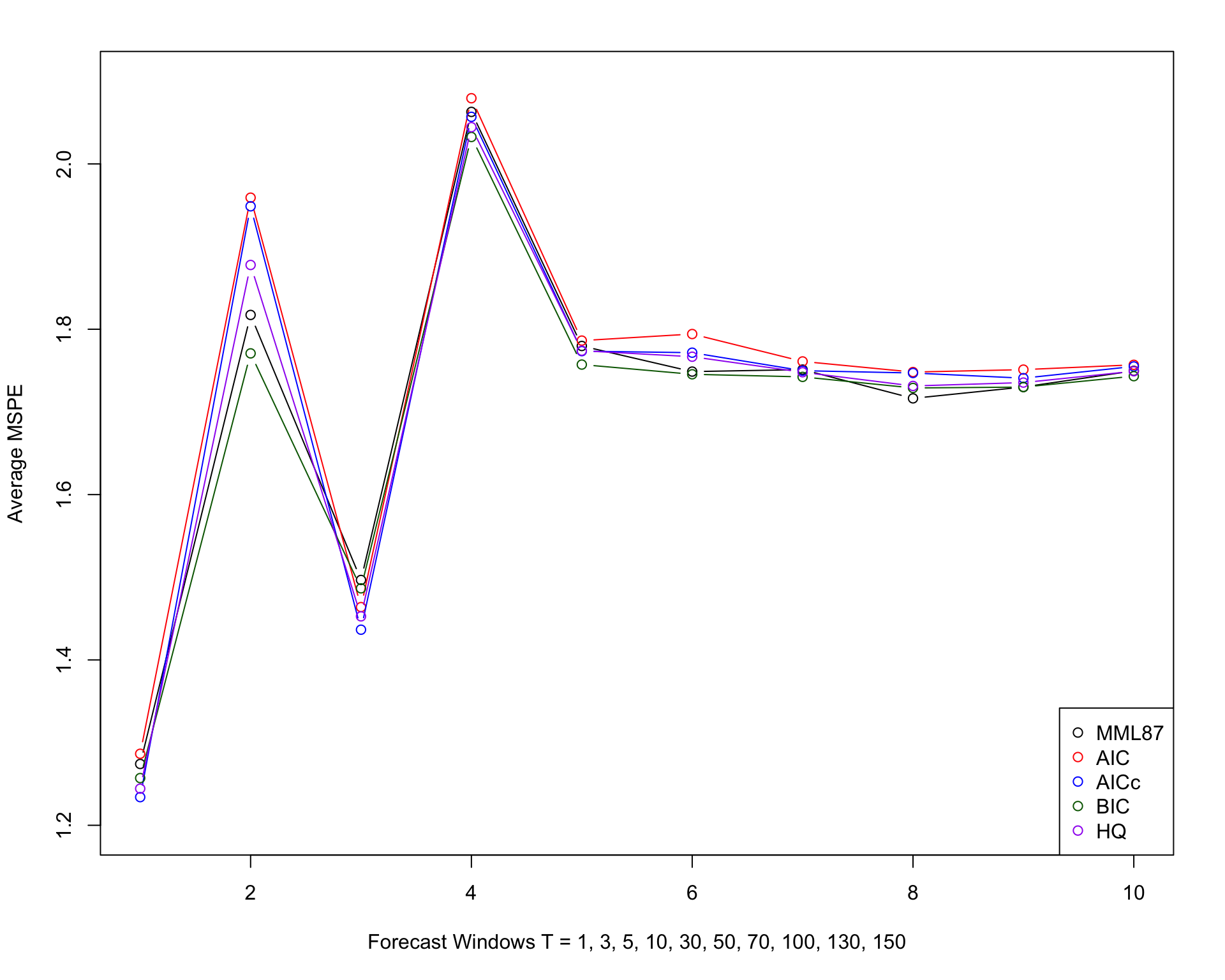}
					\caption{Average MSPE in ARMA(3, 2) Simulated Data}
					\label{arma32}
			\end{figure}

\begin{figure}[H]
				\includegraphics[width=4.88in, height=2.88in]{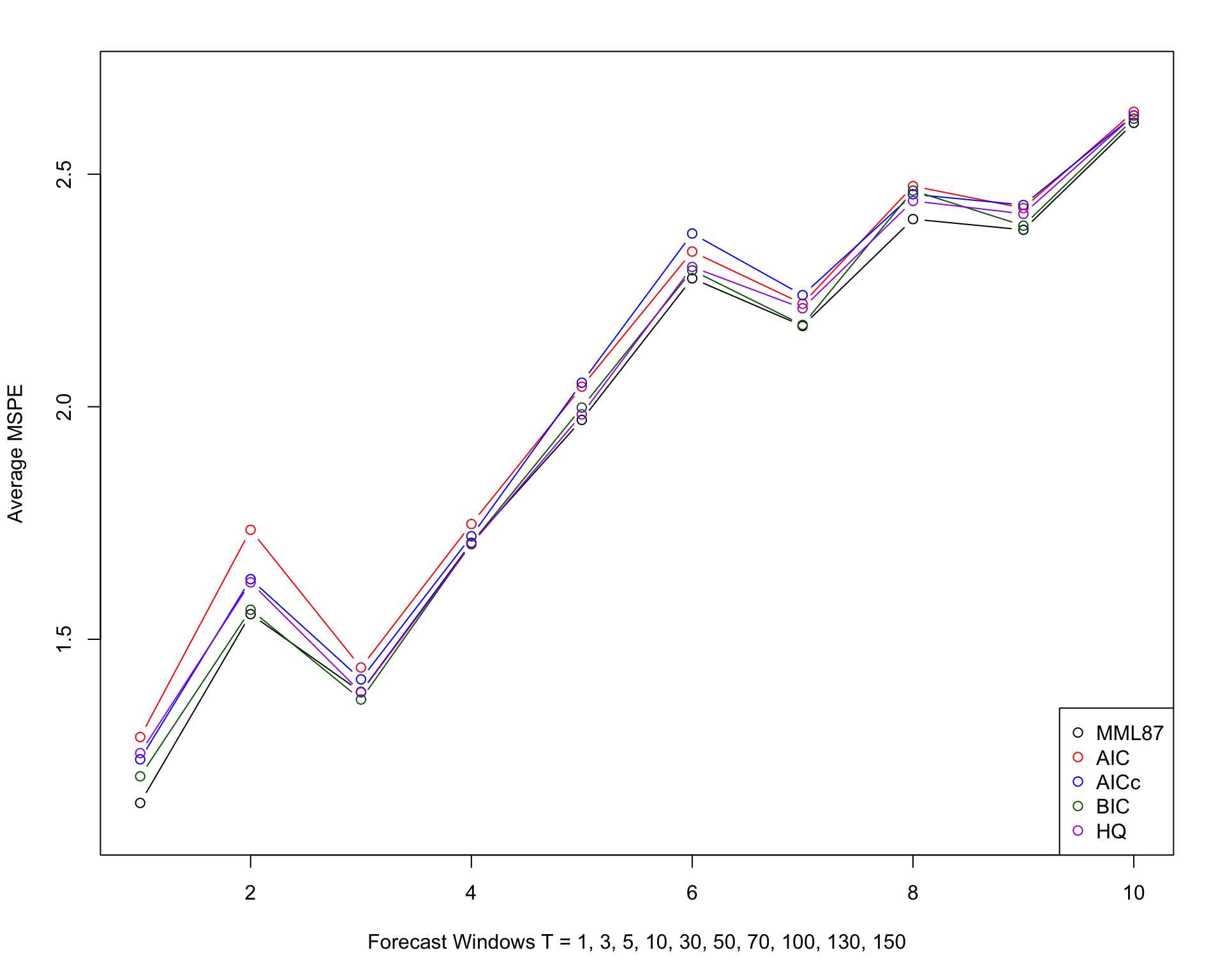}
					\caption{Average MSPE in ARMA(4, 1) Simulated Data}
					\label{arma41}
			\end{figure}

\begin{figure}[H]
				\includegraphics[width=4.88in, height=2.88in]{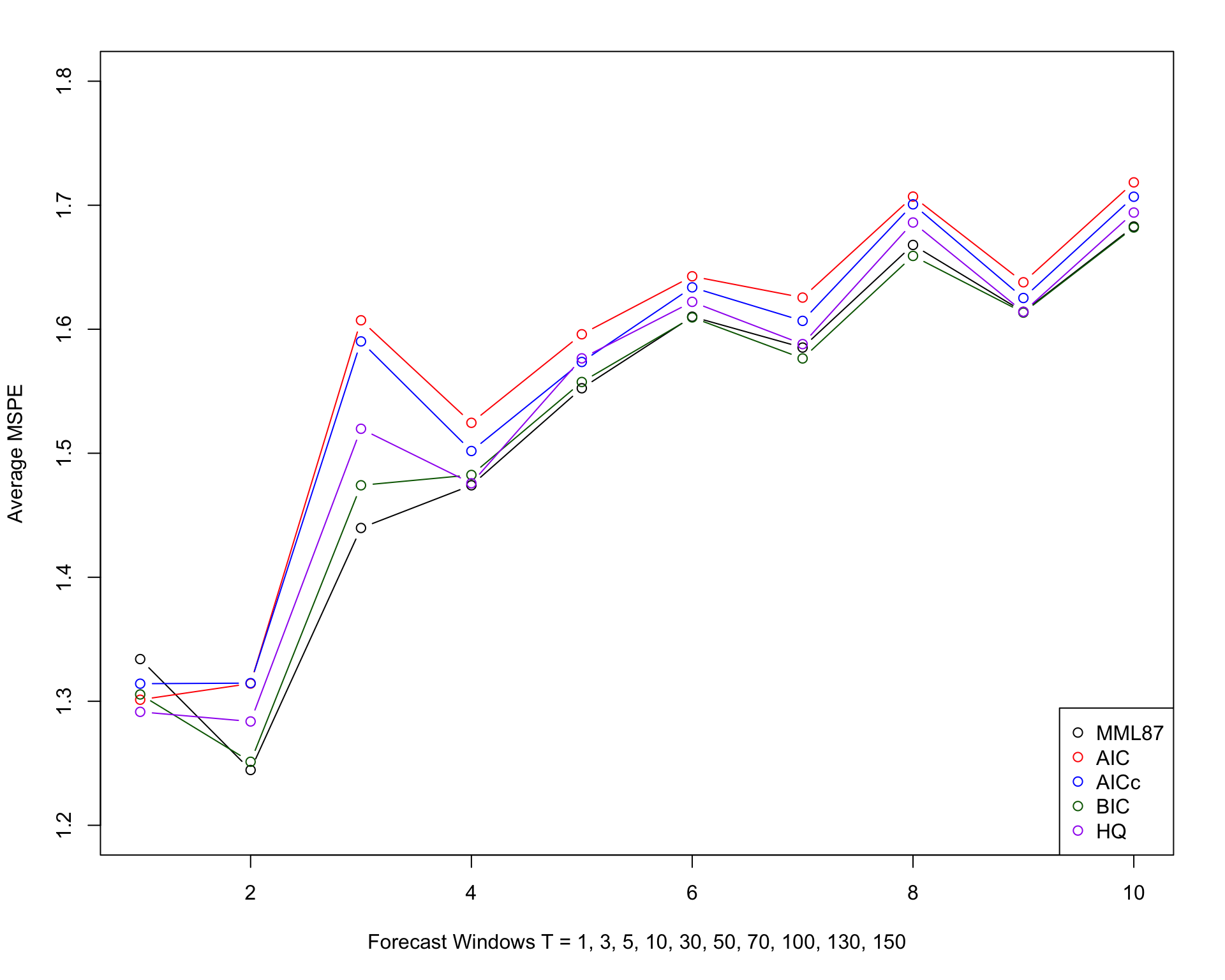}
					\caption{Average MSPE in ARMA(4, 2) Simulated Data}
					\label{arma42}
			\end{figure}

\begin{figure}[H]
				\includegraphics[width=4.88in, height=2.88in]{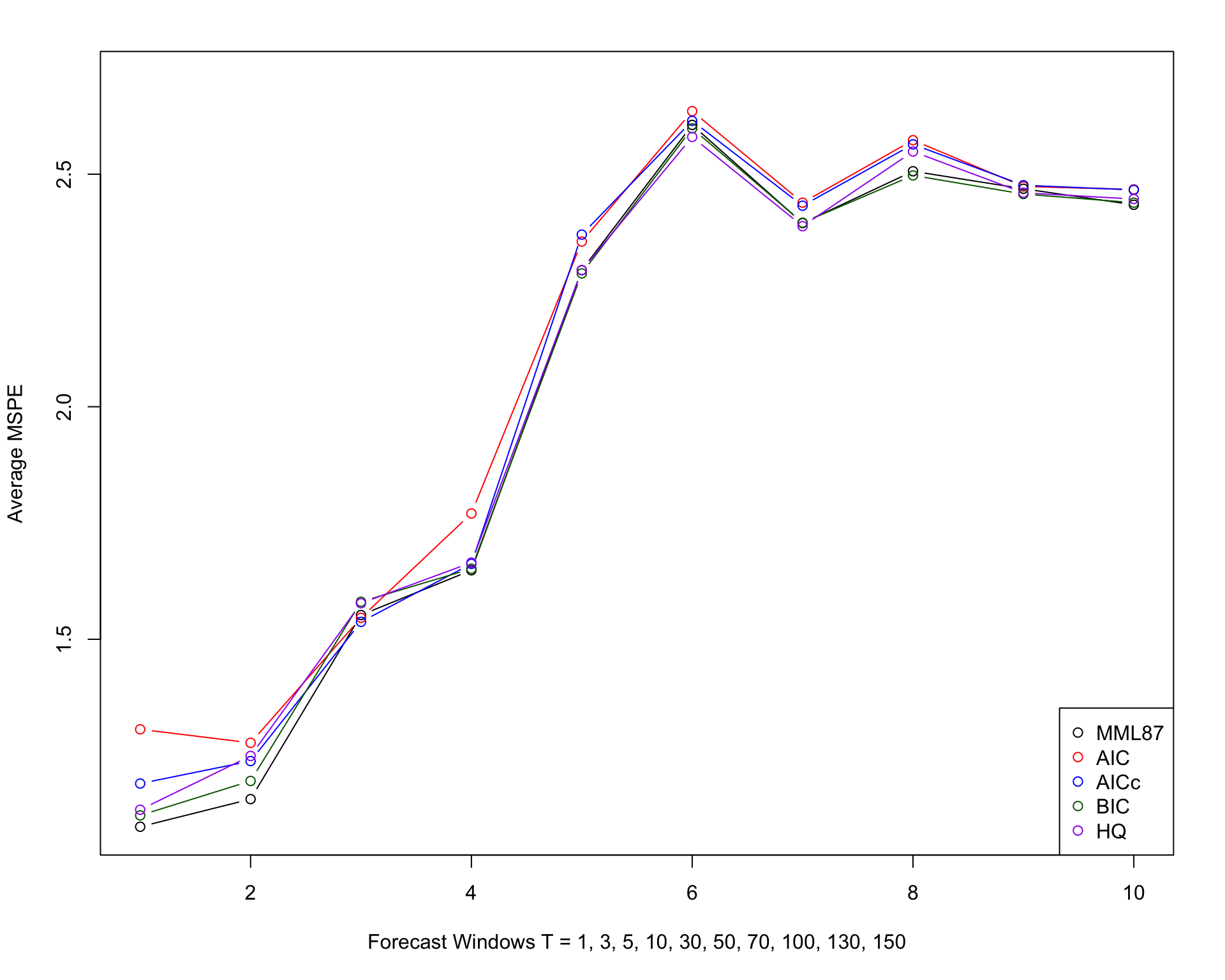}
					\caption{Average MSPE in ARMA(5, 1) Simulated Data}
					\label{arma51}
			\end{figure}

\begin{figure}[H]
				\includegraphics[width=4.88in, height=2.88in]{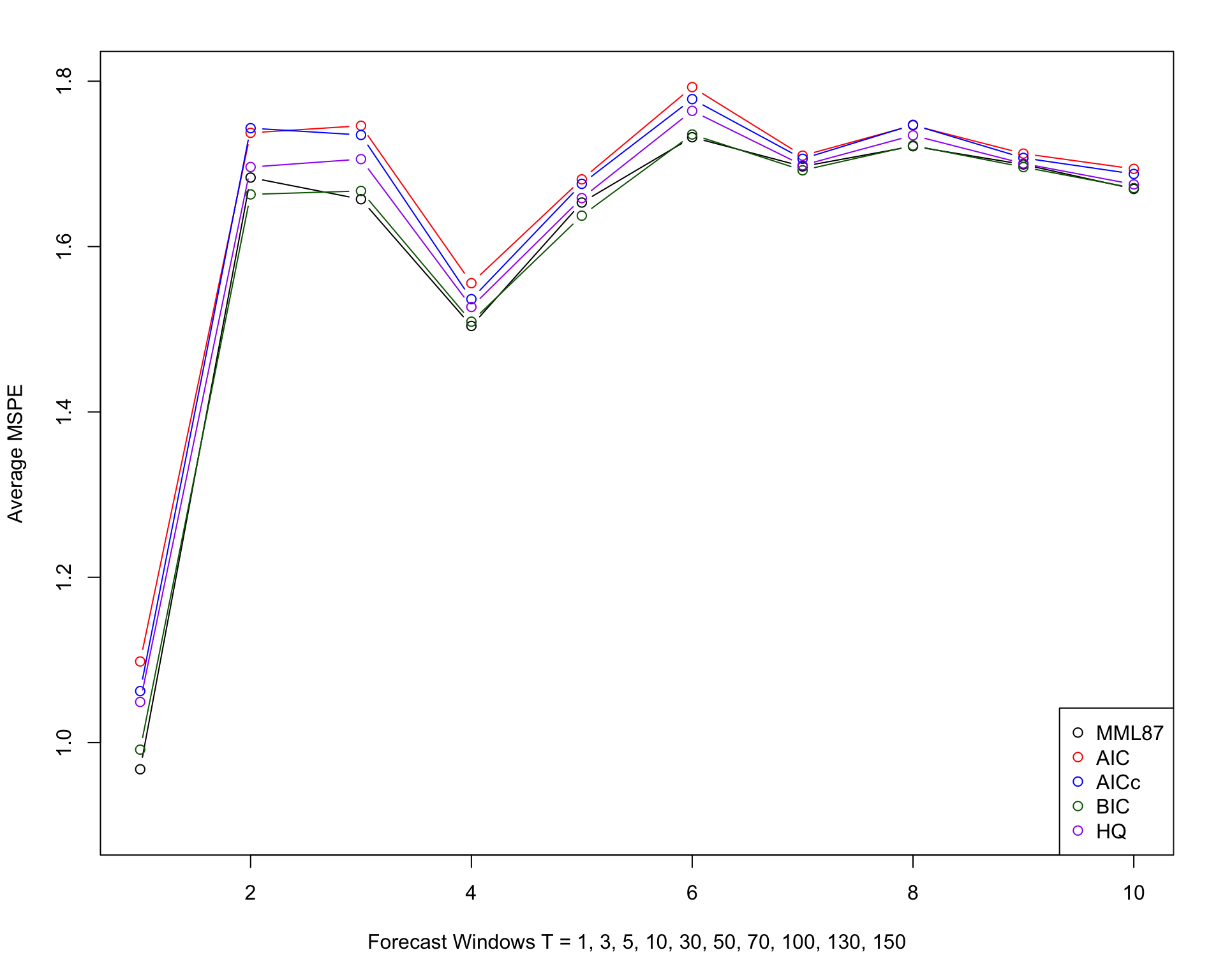}
					\caption{Average MSPE in ARMA(5, 2) Simulated Data}
					\label{arma52}
			\end{figure}

			Figure \ref{arma21} shows the average MSPE comparison between the MML87, AIC, AICc, BIC, and HQ, the simulated data generated from parameters ARMA(2, 1). The Figure 2 shows the comparison in terms of simulated data generated from ARMA(3, 3) parameters. The MML is outperform the other information criteria because the mean of prediction error is the smallest across the different forecast window sizes. Figures from \ref{arma22} to \ref{arma52} in section \ref{appendixfigure} show the diagram in the data generated from parameters ARMA(2, 2) to ARMA(5, 2).
			
		\subsection{Actual data}
			In this section, we are using the real financial data with time series characteristics to compare the performance for AIC, AICc, BIC, HQ, and MML87. In considering both the systematic and unsystematic risk, the data collected from the stock index from the market portfolio across different countries including the ASX200, Dow Jones Composite Average, FTSE100, Russell, Nasdaq 100, Nasdaq Composite, and NYSE Composite, also includes two individual stock AAPL and GS. The time horizon that the financial data selected is from 2020-09-08 to 2021-09-07. Table \ref{countingfinancialdata} shows the number of times that the information criteria selected the lower error model. For each asset portfolio, the data are separated into the 8 segments with N = 30 and T = 10, and use different combinations of lag order 1 to 5 in AR$(p)$ and lag order 0 to 5 in MA$(q)$. The results suggest that the MML87 is able to select a larger number of lower error models than other information criteria in 8 assets out of 10, the best information criterion shown in bold text.
			\begin{table}[H]
			    \centering
				\begin{tabular}{|c|c|c|c|c|c|}
					\hline
					Criterions & MML87 & AIC & AICc & BIC & HQ \\ 
					\hline
					Dow Jones Composite Average & \textbf{4} & 3 & \textbf{4} & 3 & \textbf{4} \\
					\hline
					SP500 & \textbf{5} & 2 & \textbf{5}\textbf{} & \textbf{5} & 4 \\
					\hline
					ASX200 & \textbf{7} & 3 & 5 & 5 & 4 \\
					\hline
					Russell & 2 & \textbf{5} & \textbf{5} & \textbf{5} & \textbf{5} \\
					\hline
					Dow Jones Industrial Average & 4 & 4 & \textbf{6} & \textbf{6} & 3 \\
					\hline
					FTSE100 & \textbf{5} & \textbf{5} & 4 & 4 & \textbf{5} \\
					\hline
					Nasdaq100 & \textbf{7} & 3 & 2 & 3 & 3 \\
					\hline
					Nasdaq Composite & \textbf{5} & 3 & 2 & 3 & 2 \\
					\hline
					AAPL  & \textbf{5} & \textbf{5} & 4 & \textbf{5} & 3 \\
					\hline
					GS & \textbf{6} & 4 & 4 & 4 & 4 \\
					\hline
				\end{tabular}
				\caption{\centering Number of times that information-theoretic selects the minimum forecast error in financial data}
				\label{countingfinancialdata}
			\end{table}
			
			Because the selected portfolio values have different data scales, to avoid the misclassification of best model selection criteria. Table \ref{errorfinancialdata} provides the average of the log prediction error for ten different asset portfolios. The bold texts are the minimum error in this asset portfolio. The results show the models selected by MML87 beat the AIC, AICc, BIC, and HQ with having a lower average of log prediction error.
			\begin{table}[H]
			    \centering
				\begin{tabular}{|c|c|c|c|c|c|}
					\hline
					 & MML87 & AIC & AICc & BIC & HQ \\ 
					\hline
					DJ Com  & \multicolumn{1}{|p{1.5cm}|}{\centering 9.339222 \\ (1.48159)} & \multicolumn{1}{|p{1.5cm}|}{\centering 9.431943 \\ (1.48159)} & \multicolumn{1}{|p{1.5cm}|}{\centering \textbf{9.266359} \\ (1.50159)} & \multicolumn{1}{|p{1.5cm}|}{\centering 9.350298 \\ (1.43139)} & \multicolumn{1}{|p{1.5cm}|}{\centering 9.266359 \\ (1.39585)} \\ 
					\hline
                    SP500  & \multicolumn{1}{|p{1.5cm}|}{\centering \textbf{8.170703} \\ (0.95529)} & \multicolumn{1}{|p{1.5cm}|}{\centering 8.467639 \\ (0.95529)} & \multicolumn{1}{|p{1.5cm}|}{\centering 8.175019 \\ (1.19616)} & \multicolumn{1}{|p{1.5cm}|}{\centering 8.175019 \\ (0.94133)} & \multicolumn{1}{|p{1.5cm}|}{\centering 8.209827 \\ (0.89824)} \\ 
                    \hline
                    ASX200  & \multicolumn{1}{|p{1.5cm}|}{\centering \textbf{8.668513} \\ (0.94712)} & \multicolumn{1}{|p{1.5cm}|}{\centering 9.073504 \\ (0.94712)} & \multicolumn{1}{|p{1.5cm}|}{\centering 8.807688 \\ (1.04215)} & \multicolumn{1}{|p{1.5cm}|}{\centering 8.807688 \\ (0.91327)} & \multicolumn{1}{|p{1.5cm}|}{\centering 8.807688 \\ (0.91327)} \\ 
                    \hline
                    Russell  & \multicolumn{1}{|p{1.5cm}|}{\centering 7.556093 \\ (0.97889)} & \multicolumn{1}{|p{1.5cm}|}{\centering 7.464931 \\ (0.97889)} & \multicolumn{1}{|p{1.5cm}|}{\centering \textbf{7.362797} \\ (1.27475)} & \multicolumn{1}{|p{1.5cm}|}{\centering \textbf{7.362797} \\ (1.13518)} & \multicolumn{1}{|p{1.5cm}|}{\centering \textbf{7.362797} \\ (1.13518)} \\ 
                    \hline
                    Dow Jones  & \multicolumn{1}{|p{1.5cm}|}{\centering 12.124174 \\ (1.28175)} & \multicolumn{1}{|p{1.5cm}|}{\centering 12.279021 \\ (1.28175)} & \multicolumn{1}{|p{1.5cm}|}{\centering \textbf{11.930101} \\ (1.49217)} & \multicolumn{1}{|p{1.5cm}|}{\centering \textbf{11.930101} \\ (1.39267)} & \multicolumn{1}{|p{1.5cm}|}{\centering 12.066458 \\ (1.45273)} \\ 
                    \hline
                    FTSE100  & \multicolumn{1}{|p{1.5cm}|}{\centering \textbf{9.610056} \\ (0.92881)} & \multicolumn{1}{|p{1.5cm}|}{\centering 9.732419 \\ (0.92881)} & \multicolumn{1}{|p{1.5cm}|}{\centering 9.740016 \\ (0.87461)} & \multicolumn{1}{|p{1.5cm}|}{\centering 9.740016 \\ (0.88376)} & \multicolumn{1}{|p{1.5cm}|}{\centering 9.719694 \\ (0.90703)} \\
                    \hline
                    Nasdaq 100  & \multicolumn{1}{|p{1.5cm}|}{\centering \textbf{11.247209} \\ (0.88016)} & \multicolumn{1}{|p{1.5cm}|}{\centering 11.297879 \\ (0.88016)} & \multicolumn{1}{|p{1.5cm}|}{\centering 11.30298 \\ (0.73106)} & \multicolumn{1}{|p{1.5cm}|}{\centering 11.253671 \\ (0.76663)} & \multicolumn{1}{|p{1.5cm}|}{\centering 11.300069 \\ (0.73563)} \\ 
                    \hline
                    Nasdaq Com  & \multicolumn{1}{|p{1.5cm}|}{\centering \textbf{11.172626} \\ (0.97813)} & \multicolumn{1}{|p{1.5cm}|}{\centering 11.494775 \\ (0.97813)} & \multicolumn{1}{|p{1.5cm}|}{\centering 11.261841 \\ (0.90478)} & \multicolumn{1}{|p{1.5cm}|}{\centering 11.201055 \\ (0.9338)} & \multicolumn{1}{|p{1.5cm}|}{\centering 11.37422 \\ (0.91929)} \\ 
                    \hline
                    AAPL & \multicolumn{1}{|p{1.5cm}|}{\centering 3.082 \\ (1.21437)} & \multicolumn{1}{|p{1.5cm}|}{\centering 2.967387 \\ (1.21437)} & \multicolumn{1}{|p{1.5cm}|}{\centering 3.053346 \\ (1.22936)} & \multicolumn{1}{|p{1.5cm}|}{\centering \textbf{2.871599} \\ (1.08716)} & \multicolumn{1}{|p{1.5cm}|}{\centering 3.132657 \\ (1.10378)} \\ 
                    \hline
                    GS & \multicolumn{1}{|p{1.5cm}|}{\centering \textbf{4.714533} \\ (0.82053)} & \multicolumn{1}{|p{1.5cm}|}{\centering 4.931766 \\ (0.82053)} & \multicolumn{1}{|p{1.5cm}|}{\centering 4.776378 \\ (0.68177)} & \multicolumn{1}{|p{1.5cm}|}{\centering 4.776378 \\ (0.89269)} & \multicolumn{1}{|p{1.5cm}|}{\centering 4.786212 \\ (0.88672)} \\
                    \hline
				\end{tabular}
				\caption{\centering Average of MSPE for the model selected by different information-theoretic in financial data}
				\label{errorfinancialdata}
			\end{table}
			
	\section{Conclusion}
		We have investigated the Autoregressive–Moving-Average model in the MML87 information criteria based on the Wallace and Freeman (1987) approximation. Using the maximum likelihood estimate in ARMA modeling, and unconditional likelihood function in calculation of Fisher Information matrix. The results show MML87 outperformance but not dominate the other information criteria in the simulated data and actual financial data when using the MSPE$(p)$. On average, MML87 is a really good model selection technique in time series data.
		
	\section{Appendix}
	    \subsection{Tables}
    	    \begin{table}[H]
    			    \centering
    			    \begin{tabular}{|c|c|c|c|c|c|}
    			        \hline
    					 & MML87 & AIC & AICc & BIC & HQ \\ 
    					\hline
    					$\phi_1, \theta_1$ & \textbf{48} & 35 & 42 & \textbf{48} & 42 \\
    					\hline
    					$\phi_1, \theta_2$ & 61 & 65 & 66 & 65 & \textbf{67} \\
    					\hline
    					$\phi_2, \theta_1$ & 33 & 39 & 41 & \textbf{54} & 48 \\
    					\hline
    					$\phi_2, \theta_2$ & \textbf{66} & 48 & 57 & \textbf{66} & 58 \\
    					\hline
    					$\phi_3, \theta_1$ & 40 & 40 & 47 & \textbf{53} & 52 \\
    					\hline
    					$\phi_3, \theta_2$ & \textbf{66} & 58 & 63 & 63 & 64 \\
    					\hline
    					$\phi_4, \theta_1$ & 30 & 47 & 50 & \textbf{57} & \textbf{54} \\
    					\hline
    					$\phi_4, \theta_2$ & \textbf{69} & 61 & 59 & \textbf{69} & 60 \\
    					\hline
    					$\phi_5, \theta_1$ & \textbf{53} & 41 & 39 & 36 & 40 \\
    					\hline
    					$\phi_5, \theta_2$ & 49 & 47 & 54 & \textbf{64} & 57 \\
    					\hline
    			    \end{tabular}
    			    \caption{\centering Number of times that information-theoretic selects the minimum forecast error in N = 50 T = 10}
    			    \label{countingcompare50}
    			\end{table}
    			
    			\begin{table}[H]
    			    \centering
    			    \begin{tabular}{|c|c|c|c|c|c|}
    			        \hline
    					 & MML87 & AIC & AICc & BIC & HQ \\ 
    					\hline
    					$\phi_1, \theta_1$ & \textbf{48} & 40 & 43 & 41 & 40 \\
    					\hline
    					$\phi_1, \theta_2$ & \textbf{75} & 64 & 65 & 67 & 67 \\
    					\hline
    					$\phi_2, \theta_1$ & \textbf{43} & 37 & 34 & \textbf{43} & 39 \\
    					\hline
    					$\phi_2, \theta_2$ & 66 & 62 & 61 & \textbf{71} & 68 \\
    					\hline
    					$\phi_3, \theta_1$ & \textbf{48} & 37 & 41 & 31 & 40 \\
    					\hline
    					$\phi_3, \theta_2$ & \textbf{74} & 60 & 63 & 73 & 73 \\
    					\hline
    					$\phi_4, \theta_1$ & 41 & 46 & \textbf{50} & 43 & 43 \\
    					\hline
    					$\phi_4, \theta_2$ & 71 & 60 & 60 & \textbf{77} & 73 \\
    					\hline
    					$\phi_5, \theta_1$ & 45 & 46 & 44 & \textbf{49} & 47 \\
    					\hline
    					$\phi_5, \theta_2$ & \textbf{67} & 57 & 58 & 66 & 61 \\
    					\hline
    			    \end{tabular}
    			    \caption{\centering Number of times that information-theoretic selects the minimum forecast error in N = 150 T = 10}
    			    \label{countingcompare150}
    			\end{table}
    			
    			\begin{table}[H]
    			    \centering
    			    \begin{tabular}{|c|c|c|c|c|c|}
    			        \hline
    					 & MML87 & AIC & AICc & BIC & HQ \\ 
    					\hline
    					$\phi_1, \theta_1$ & \textbf{47} & 33 & 37 & 36 & 36 \\
    					\hline
    					$\phi_1, \theta_2$ & \textbf{65} & 57 & 54 & 64 & 57 \\
    					\hline
    					$\phi_2, \theta_1$ & \textbf{55} & 51 & 49 & 44 & 40 \\
    					\hline
    					$\phi_2, \theta_2$ & \textbf{68} & 57 & 59 & \textbf{69} & 59 \\
    					\hline
    					$\phi_3, \theta_1$ & 42 & 47 & \textbf{50} & 47 & \textbf{50} \\
    					\hline
    					$\phi_3, \theta_2$ & 68 & 59 & 59 & \textbf{71} & 62 \\
    					\hline
    					$\phi_4, \theta_1$ & 52 & \textbf{53} & 49 & 49 & 49 \\
    					\hline
    					$\phi_4, \theta_2$ & \textbf{71} & 62 & 62 & 68 & 63 \\
    					\hline
    					$\phi_5, \theta_1$ & 51 & 51 & 49 & 53 & \textbf{55} \\
    					\hline
    					$\phi_5, \theta_2$ & 39 & \textbf{55} & 52 & 43 & 50 \\
    					\hline
    			    \end{tabular}
    			    \caption{\centering Number of times that information-theoretic selects the minimum forecast error in N = 200 T = 10}
    			    \label{countingcompare200}
    			\end{table}
    			
    			\begin{table}[H]
    			    \centering
    			    \begin{tabular}{|c|c|c|c|c|c|}
    			        \hline
    					 & MML87 & AIC & AICc & BIC & HQ \\ 
    					\hline
    					$\phi_1, \theta_1$ & 30 & 44 & \textbf{45} & \textbf{45} & 43 \\
    					\hline
    					$\phi_1, \theta_2$ & \textbf{75} & 49 & 52 & 67 & 65 \\
    					\hline
    					$\phi_2, \theta_1$ & \textbf{56} & 50 & 49 & 51 & 53 \\
    					\hline
    					$\phi_2, \theta_2$ & 64 & 69 & \textbf{70} & 67 & 68 \\
    					\hline
    					$\phi_3, \theta_1$ & 32 & 49 & \textbf{51} & 29 & 34 \\
    					\hline
    					$\phi_3, \theta_2$ & \textbf{65} & 61 & 60 & 62 & 59 \\
    					\hline
    					$\phi_4, \theta_1$ & 55 & 46 & 48 & 55 & \textbf{59} \\
    					\hline
    					$\phi_4, \theta_2$ & \textbf{72} & 61 & 63 & \textbf{72} & 69 \\
    					\hline
    					$\phi_5, \theta_1$ & \textbf{61} & 48 & 49 & 44 & 44 \\
    					\hline
    					$\phi_5, \theta_2$ & 50 & 44 & 44 & \textbf{59} & 55 \\
    					\hline
    			    \end{tabular}
    			    \caption{\centering Number of times that information-theoretic selects the minimum forecast error in N = 300 T = 10}
    			    \label{countingcompare300}
    			\end{table}
    			
    			\begin{table}[H]
    			    \centering
    			    \begin{tabular}{|c|c|c|c|c|c|}
    			        \hline
    					 & MML87 & AIC & AICc & BIC & HQ \\ 
    					\hline
    					$\phi_1, \theta_1$ & 47 & 42 & 41 & \textbf{49} & 38 \\
    					\hline
    					$\phi_1, \theta_2$ & \textbf{74} & 71 & 73 & 73 & \textbf{74} \\
    					\hline
    					$\phi_2, \theta_1$ & 42 & 48 & \textbf{50} & 44 & 46 \\
    					\hline
    					$\phi_2, \theta_2$ & \textbf{73} & 62 & 60 & \textbf{73} & \textbf{73} \\
    					\hline
    					$\phi_3, \theta_1$ & \textbf{45} & 44 & 41 & 39 & 40 \\
    					\hline
    					$\phi_3, \theta_2$ & 67 & 60 & 60 & 65 & \textbf{69} \\
    					\hline
    					$\phi_4, \theta_1$ & 38 & 47 & 47 & \textbf{50} & 45 \\
    					\hline
    					$\phi_4, \theta_2$ & \textbf{70} & 51 & 51 & \textbf{70} & \textbf{70} \\
    					\hline
    					$\phi_5, \theta_1$ & 54 & \textbf{57} & 55 & 50 & 51 \\
    					\hline
    					$\phi_5, \theta_2$ & \textbf{65} & 58 & 60 & \textbf{65} & 60 \\
    					\hline
    			    \end{tabular}
    			    \caption{\centering Number of times that information-theoretic selects the minimum forecast error in N = 500 T = 10}
    			    \label{countingcompare500}
    			\end{table}
    			
    	\subsection{Figures}
    	\label{appendixfigure}
    	    
    \printbibliography[title={Reference}]

@inproceedings{fitzgibbon2004minimum,
  title={Minimum message length autoregressive model order selection},
  author={Fitzgibbon, Leigh J and Dowe, David L and Vahid, Farshid},
  booktitle={International Conference on Intelligent Sensing and Information Processing, 2004. Proceedings of},
  pages={439--444},
  year={2004},
  organization={IEEE}
}

@inproceedings{sak2005minimum,
  title={Minimum message length moving average time series data mining},
  author={Sak, Mony and Dowe, David L and Ray, Sid},
  booktitle={2005 ICSC Congress on Computational Intelligence Methods and Applications},
  pages={6--pp},
  year={2005},
  organization={IEEE}
}

@article{yao2006gaussiani,
  title={Gaussian maximum likelihood estimation for ARMA models. I. time series},
  author={Yao, Qiwei and Brockwell, Peter J},
  journal={Journal of time series analysis},
  volume={27},
  number={6},
  pages={857--875},
  year={2006},
  publisher={Wiley Online Library}
}

@article{yao2006gaussianii,
  title={Gaussian maximum likelihood estimation for ARMA models II: spatial processes},
  author={Yao, Qiwei and Brockwell, Peter J},
  journal={Bernoulli},
  volume={12},
  number={3},
  pages={403--429},
  year={2006},
  publisher={Bernoulli Society for Mathematical Statistics and Probability}
}

@article{anderson1976inverse,
  title={On the inverse of the autocovariance matrix for a general moving average process},
  author={OD, Anderson},
  journal={Biometrika},
  volume={63},
  number={2},
  pages={391--394},
  year={1976},
  publisher={Oxford University Press}
}

@article{ljung1979likelihood,
  title={The likelihood function of stationary autoregressive-moving average models},
  author={Ljung, Greta M and Box, George EP},
  journal={Biometrika},
  volume={66},
  number={2},
  pages={265--270},
  year={1979},
  publisher={Oxford University Press}
}

@article{miller1995exact,
  title={Exact maximum likelihood estimation in autoregressive processes},
  author={Miller, James W},
  journal={Journal of Time Series Analysis},
  volume={16},
  number={6},
  pages={607--615},
  year={1995},
  publisher={Wiley Online Library}
}

@article{wincek1986exact,
  title={An exact maximum likelihood estimation procedure for regression-ARMA time series models with possibly nonconsecutive data},
  author={Wincek, Michael A and Reinsel, Gregory C},
  journal={Journal of the Royal Statistical Society: Series B (Methodological)},
  volume={48},
  number={3},
  pages={303--313},
  year={1986},
  publisher={Wiley Online Library}
}

@article{shephard1997relationship,
  title={The relationship between the conditional sum of squares and the exact likelihood for autoregressive moving average model.},
  author={Shephard, Neil},
  year={1997}
}

@article{klein1995computation,
  title={Computation of the Fisher information matrix for time series models},
  author={Klein, Andr{\'e} and Melard, Guy},
  journal={Journal of computational and applied mathematics},
  volume={64},
  number={1-2},
  pages={57--68},
  year={1995},
  publisher={Elsevier}
}

@misc{box2015time,
  title={Time series analysis, control, and forecasting . Hoboken},
  author={Box, GE and Jenkins, GM and Reinsel, GC and Ljung, GM},
  year={2015},
  publisher={New Jersey: John Wiley \& Sons}
}

@misc{box1976time,
  title={Time series analysis prediction and control},
  author={Box, George EP and Jenkins, Gwilym M and Reinsel, Gregory C},
  year={1976},
  publisher={Holden-Day, San Francisco}
}

@article{vahid1999partial,
  title={Partial Pooling: A Possible Answer to},
  author={Vahid, Farshid},
  journal={Cointegration, Causality, and Forecasting: A Festschrift in Honour of Clive WJ Granger},
  pages={410},
  year={1999},
  publisher={Oxford University Press on Demand}
}

@article{wallace1999minimum,
  title={Minimum message length and Kolmogorov complexity},
  author={Wallace, Chris S. and Dowe, David L.},
  journal={The Computer Journal},
  volume={42},
  number={4},
  pages={270--283},
  year={1999},
  publisher={Oxford University Press}
}

@article{ward2008review,
  title={A review and comparison of four commonly used Bayesian and maximum likelihood model selection tools},
  author={Ward, Eric J},
  journal={Ecological Modelling},
  volume={211},
  number={1-2},
  pages={1--10},
  year={2008},
  publisher={Elsevier}
}

@article{barndorff1973parametrization,
  title={On the parametrization of autoregressive models by partial autocorrelations},
  author={Barndorff-Nielsen, Ole and Schou, Geert},
  journal={Journal of multivariate Analysis},
  volume={3},
  number={4},
  pages={408--419},
  year={1973},
  publisher={Elsevier}
}

@article{piccolo1982size,
  title={The size of the stationarity and invertibility region of an autoregressive-moving average process},
  author={Piccolo, Domenico},
  journal={Journal of Time Series Analysis},
  volume={3},
  number={4},
  pages={245--247},
  year={1982},
  publisher={Wiley Online Library}
}

@article{chatfield2004cross,
  title={Cross-covariance and cross-correlation},
  author={Chatfield, C},
  journal={The Analysis of Time Series: An Introduction. Boca Raton: Chapman \& Hall/CRC},
  pages={155--159},
  year={2004}
}

@article{wallace1987estimation,
  title={Estimation and inference by compact coding},
  author={Wallace, Chris S and Freeman, Peter R},
  journal={Journal of the Royal Statistical Society: Series B (Methodological)},
  volume={49},
  number={3},
  pages={240--252},
  year={1987},
  publisher={Wiley Online Library}
}

@article{wallace1968information,
  title={An information measure for classification},
  author={Wallace, Chris S and Boulton, David M},
  journal={The Computer Journal},
  volume={11},
  number={2},
  pages={185--194},
  year={1968},
  publisher={The British Computer Society}
}

@phdthesis{wong2019minimum,
  title={Minimum message length inference with application to genome-wide association studies data},
  author={Wong, Chi Kuen},
  year={2019}
}

@article{fang2021minimum,
  title={Minimum Message Length in Hybrid ARMA and LSTM Model Forecasting},
  author={Fang, Zheng and Dowe, David L and Peiris, Shelton and Rosadi, Dedi},
  year={2021},
  publisher={Preprints}
}

@article{fang2021climate,
  title={Climate Finance: Mapping Air Pollution and Finance Market in Time Series},
  author={Fang, Zheng and Xie, Jianying and Peng, Ruiming and Wang, Sheng},
  journal={Econometrics},
  volume={9},
  number={4},
  pages={43},
  year={2021},
  publisher={Multidisciplinary Digital Publishing Institute}
}

\end{document}